\documentclass[11pt,a4paper]{article}
\usepackage{cite}

\usepackage{amsmath, amsthm,mathtools, url,amssymb,upgreek,slashed}
\usepackage{ifpdf}
\ifpdf
  \usepackage[pdftex]{graphicx}
  \usepackage{epstopdf}
\else
  \usepackage[dvips]{graphicx}
\fi
\parskip=\baselineskip

\numberwithin{equation}{section}

\DeclareMathOperator{\res}{Res}

\renewcommand{\phi}{\varphi}

\newcommand{\beq}{\begin{equation}}
\newcommand{\eeq}{\end{equation}}

\newcommand{\CC}{\mathbb{C}}
\newcommand{\RR}{\mathbb{R}}  

\newcommand{\SL}{\mathrm{SL}}

\usepackage[symbol]{footmisc}

\renewcommand{\thefootnote}{\fnsymbol{footnote}}

\begin{document}

\begin{titlepage}

\begin{flushright}
\end{flushright}

\vskip 1.5in

\begin{center}
{\bf\Large{Einstein--Rosen Waves and the Geroch Group}}

\vskip 0.5cm {Robert F. Penna\footnote[1]{rpenna@ias.edu} } 
\vskip 0.05in {\small{ \textit{Institute for Advanced Study}
\vskip -.4cm
{\textit{Einstein Drive, Princeton, NJ 08540 USA}}}
}
\end{center}
\vskip 0.5in
\baselineskip 16pt

\begin{abstract}  

Under the action of the Geroch group, the Minkowski metric can be transformed into any vacuum metric with two commuting Killing vectors.  In principle, this reduces the problem of deriving vacuum metrics with two commuting Killing vectors to pure algebra.  In this short note, we use these facts to give a purely algebraic derivation of the Einstein--Rosen metric, which describes a cylindrical gravitational wave.  Our derivation has a straightforward extension to gravitational pulse waves.

\end{abstract}


\end{titlepage}

\renewcommand*{\thefootnote}{\arabic{footnote}}
\setcounter{footnote}{0}


\section{Introduction}

The equations of motion of general relativity are a set of ten coupled nonlinear partial differential equations (PDEs).  One does not usually expect nonlinear PDEs to admit exact solutions.  But 
general relativity admits many exact solutions  \cite{bivcak1999selected,stephani2009exact,griffiths2009exact}.  It is natural ask if there is a hidden symmetry that explains why these exact solutions exist.

Integrable systems are nonlinear PDEs with very large hidden symmetry groups.  Roughly speaking, the number of symmetries equals the number of degrees of freedom\footnote{It is difficult to make this statement precise in field theory because the number of degrees of freedom is infinite, but it is a useful heuristic description of what it means to be integrable.}.  The large amount of hidden symmetry causes integrable systems to behave as if they are ``secretly linear.''  They admit exact solutions despite being nonlinear differential equations.

General relativity is not an integrable system in four spacetime dimensions.  However, it becomes an integrable system after dimensional reduction to two dimensions \cite{geroch1972method,breitenlohner1987geroch,nicolai1991two}.  The 2d theory is an $\SL(2,\RR)$ coset sigma model coupled to 2d general relativity.  Solutions of 4d general relativity with two commuting Killing vectors can be realized as solutions of the 2d theory.   For example, the Schwarzschild and Kerr black hole metrics can be realized this way.  The 2d theory has a hidden infinite dimensional  symmetry called the Geroch group which explains why it is integrable.  The existence of the Geroch group thus explains why 4d general relativity admits exact solutions with two commuting Killing vectors.

Under the action of the Geroch group, the Minkowski metric can be transformed into any vacuum metric with two commuting Killing vectors.  In principle, this reduces the problem of finding vacuum metrics with two commuting Killing vectors to pure algebra.  One does not need to solve the equations of motion, or any  PDEs, directly.  In practice, the algebra involved can become quite difficult.  An algebraic derivation of the Kerr metric only appeared in print fairly recently \cite{maison1999duality,katsimpouri2013inverse}.  
 
The Einstein--Rosen wave is an interesting example of a metric with two commuting Killing vectors that is not stationary \cite{beck1925theorie,einstein1937gravitational}.  It describes a cylindrical gravitational wave.  Historically, it played an important role in early attempts at defining the energy carried by gravitational waves \cite{thorne1965energy,chandrasekhar1986cylindrical}.  We refer to \cite{ashtekar1997asymptotic,ashtekar1997behavior} for the modern definition of the energy carried by cylindrical gravitational waves, which is based on asymptotic symmetries at null infinity.

In this short note, we give a purely algebraic derivation of the Einstein--Rosen metric using the Geroch group.  The solution requires some pleasant identities involving infinite sums of Bessel functions and squares of Bessel functions, so it is not quite trivial.  On the other hand, it is simple enough to be useful as a way to clearly see the integrable structures underlying dimensionally reduced general relativity ``in action.''

\section{The Einstein--Rosen Wave}

The Einstein--Rosen wave is an exact solution of general relativity that describes a cylindrical gravitational wave.  In Weyl's canonical coordinates, $(t,\rho,\phi,z)$, the metric is
\beq\label{eq:metric}
ds^2 = 
	e^{2\gamma-2\psi}(-dt^2 + d\rho^2) + e^{-2\psi} \rho^2d\phi^2 + e^{2\psi} dz^2  \,,
\eeq
where
\begin{align}
\psi	&= J_0(\rho) \cos t \,, \label{eq:psi} \\
\gamma &= \frac{1}{2}\rho^2 J_0(\rho)^2 + \frac{1}{2}\rho^2 J_1(\rho)^2 
		- \rho J_0(\rho) J_1(\rho) \cos^2t \,. \label{eq:gamma}
\end{align}
$J_0(\rho)$ and $J_1(\rho)$ are Bessel functions.  The range of the radial coordinate is $\rho \geq 0$ and the range of the angular coordinate is $0\leq \phi<2 \pi$.  The $t$ and $z$ coordinates run from $-\infty$ to $+\infty$.  

The wave is periodic in $t$.   It is supported on an infinitely long cylinder in the $\phi$ and $z$ directions.   There are two (spacelike) commuting Killing vectors, $\partial_\phi$ and $\partial_z$. 

We have set the wave number to $k=1$.  It can be restored by replacing $t\rightarrow k t$ and $\rho \rightarrow k\rho$ (constant $k$) in equations \eqref{eq:psi}--\eqref{eq:gamma}.  
It is possible to obtain gravitational ``pulse waves'' by taking superpositions of Einstein--Rosen waves with different wave numbers \cite{rosen1954some,bonnor1957non,weber1957reality}. 
Our results have a straightforward extension to pulse waves.

\section{The Monodromy Function}

The monodromy matrix is an invariant that can be attached to any vacuum metric with two commuting Killing vectors \cite{breitenlohner1987geroch}.  Any such metric can be fully recovered from its monodromy matrix\footnote{The prescription we use to recover the metric is based on \cite{breitenlohner1987geroch}.  Different prescriptions exist and different prescriptions may assign different metrics to the same monodromy matrix \cite{aniceto2020weyl}.}.   The Minkowski metric can be transformed into any other vacuum metric with two commuting Killing vectors by acting on its monodromy matrix with an element of the Geroch group, and then by recovering the new metric from the new monodromy matrix.  The nontrivial step is recovering the metric from the monodromy matrix.

In general, the monodromy matrix is an $\SL(2,\RR)$-valued function of a complex parameter, $\tau$, called the spectral parameter.   However,  diagonal metrics\footnote{The notion of diagonal metric is coordinate dependent.  Here and in what follows, we are assuming the metric is in Weyl's canonical coordinates.} have diagonal monodromy matrices, and a diagonal $\SL(2,\RR)$ matrix is characterized by a single number.  So a diagonal metric can be characterized by a monodromy function, $m(\tau)$.

The monodromy function of the Einstein--Rosen metric is
\beq\label{eq:monodromy}
m(\tau)  = \cos \tau \,,  \quad (\tau \in \CC)\,.
\eeq
$\tau$ is the spectral parameter.

There is a simple rule for getting the monodromy function from the metric \eqref{eq:metric}.  First, evaluate the metric function 
 $\psi(t,\rho)$ at $\rho=0$ to get $\psi(t,0) = \cos t$.  Then, complexify $t$ to get $m(\tau) = \cos \tau$, $\tau\in \CC$.  This rule applies to any diagonal vacuum metric of the form \eqref{eq:metric}.  For example, it applies to gravitational pulse waves. 

The more interesting and less straightforward part of the problem is recovering the metric from the monodromy function.  At first glance this looks hopeless, because the monodromy function \eqref{eq:monodromy} is so much simpler than the metric.  To see why things are not so grim, observe that our task is a bit like a boundary value problem, because the monodromy function is (a complexified version of) the boundary value of $\psi(t,\rho)$ at the $\rho=0$ boundary of the 2d spacetime.   The surprising part of the story is that this problem has a purely algebraic solution and we do not need to solve any PDEs.

An important property of integrable systems in curved spacetime is that they typically admit two spectral parameters.  The first one, $\tau$, is called the constant spectral parameter because $m(\tau)$ does not depend on the spacetime coordinates.  The second one, $\zeta$, is called the spacetime-dependent spectral parameter. It is related to $\tau$ by 
\beq\label{eq:tau}
\tau  = \frac{\rho}{2} (\zeta + \zeta^{-1})  - t \,.
\eeq
The best way to understand this formula is using twistors \cite{woodhouse1988geroch,penna2020twistor}.  A point in spacetime corresponds to a line in twistor space. $\zeta$ is a coordinate on the twistor line.  Surfaces with $\tau={\rm constant}$ are surfaces in twistor space that are invariant under the action of $\partial_\phi$ and $\partial_z$.  We stress that equation \eqref{eq:tau} does not depend on the details of the metric.  It applies to all spacetimes with  Killing vectors $\partial_\phi$ and $\partial_z$.  This includes gravitational pulse waves and even more general spacetimes with nondiagonal metrics.

As a function of $\zeta$, the monodromy function of the Einstein--Rosen wave is
\begin{align}
\cos\tau 
	&= \frac{e^{i\tau} + e^{-i\tau}}{2} \notag \\
	&= \frac{1}{2}e^{-it}e^{ i\frac{\rho}{2} (\zeta + \zeta^{-1}) } + \frac{1}{2}e^{it}e^{-i\frac{\rho}{2} (\zeta + \zeta^{-1}) } \label{eq:mzeta}\,.
\end{align}
This is a step closer to the metric because the spacetime coordinates, $t$ and $\rho$, have reentered the scene.

The next step is to expand \eqref{eq:mzeta} as a Laurent series in $\zeta$.   The trick is to observe that the exponentials in equation \eqref{eq:mzeta} look a lot like the Bessel function generating function,
\beq
e^{\frac{\rho}{2}(\lambda-\lambda^{-1})} = \sum_{n=-\infty}^{\infty} J_n(\rho) \lambda^n \,.
\eeq
To get an exact match, set $\lambda = \pm i\zeta$, and
\beq
e^{\pm i\frac{\rho}{2}(\zeta+\zeta^{-1})} = \sum_{n=-\infty}^{\infty} (\pm 1)^n i^n J_n(\rho) \zeta^n \,.
\eeq
Plugging into \eqref{eq:mzeta} gives the monodromy function as a Laurent series in $\zeta$:
\beq\label{eq:mseries}
\cos \tau = \frac{1}{2} e^{-it} \sum_{n=-\infty}^{\infty} i^n J_n(\rho) \zeta^n
	+ \frac{1}{2} e^{it}  \sum_{n=-\infty}^{\infty} (-1)^n i^n J_n(\rho) \zeta^n \,.
\eeq
The (re)appearance of the Bessel functions is a good sign, we are getting closer to the metric.  

Now let
\beq
\cos \tau \equiv m_+ + m_0 + m_- \,,
\eeq
where $m_+$ are the terms in \eqref{eq:mseries} with $n>0$, $m_0$ are the terms with $n=0$, and $m_-$ are the terms with $n<0$.

The terms with $n=0$ are
\beq
m_0 = \frac{1}{2} e^{-it} J_0(\rho) + \frac{1}{2}  e^{it} J_0(\rho) = J_0(\rho)\cos t \,.
\eeq
This is precisely
\beq
\psi(t,\rho) = m_0 = J_0(\rho)\cos t \,.
\eeq
Thus we recover the metric function \eqref{eq:psi} from the $n=0$ terms in the monodromy function.  This rule, $\psi(t,\rho) = m_0$, applies to all diagonal metrics.  For example, it applies to gravitational pulse waves.

The terms with $n\neq 0$ are
\begin{align}
m_+	&= \frac{1}{2}e^{-it} \sum_{n=1}^{\infty} i^n J_n(\rho) \zeta^n 
	+ \frac{1}{2} e^{it} \sum_{n=1}^{\infty} (-1)^n i^n J_n(\rho) \zeta^n \,, \\
m_-	&= \frac{1}{2}e^{-it} \sum_{m=1}^{\infty} i^m J_m(\rho) \zeta^{-m} 
	+ \frac{1}{2} e^{it} \sum_{m=1}^{\infty} (-1)^m i^m J_m(\rho) \zeta^{-m} \,.
\end{align}
In the second line, we reindexed the sums using the basic identity $J_{-m}(\rho) = (-1)^m J_m(\rho)$.

The remaining metric function \eqref{eq:gamma}, $\gamma(t,\rho)$, can be recovered from $m_\pm$ using the remarkable formula
\beq
\gamma(t,\rho) = 2\res_{\zeta = 0} \left(  m_- \frac{d}{d\zeta}  m_+ \right) . \label{eq:res}
\eeq
This is a special case of a formula of Breitenlohner and Maison \cite{breitenlohner1987geroch}.  It is related to a 2-cocycle on the $\SL(2,\RR)$ loop group.  It can be interpreted as a tau function \cite{mason2000tau,mason2002tau}.  

The derivative inside the parentheses is
\beq
\frac{d}{d\zeta} m_+ = 
	\frac{1}{2} e^{-it} \sum_{n=1}^{\infty} n i^n J_n(\rho) \zeta^{n-1} + \frac{1}{2} e^{it} \sum_{n=1}^{\infty} n (-1)^n i^n J_n(\rho) \zeta^{n-1} \,.
\eeq
Expanding the product inside the parentheses in \eqref{eq:res} gives four double infinite sums:
\begin{align}
m_- \frac{d}{d\zeta} m_+ = 
 	 & \frac{1}{4} e^{-2it} \left(\sum_{m=1}^{\infty} i^m J_m(\rho) \zeta^{-m} \right) \left( \sum_{n=1}^{\infty} n i^n J_n(\rho) \zeta^{n-1} \right) \notag \\
	 & + \frac{1}{4} \left(\sum_{m=1}^{\infty} i^m J_m(\rho) \zeta^{-m}\right)\left(\sum_{n=1}^{\infty} n (-1)^n i^n J_n(\rho) \zeta^{n-1} \right) \notag \\
 	 & + \frac{1}{4} \left(\sum_{m=1}^{\infty} (-1)^m i^m J_m(\rho) \zeta^{-m} \right)\left( \sum_{n=1}^{\infty} n i^n J_n(\rho) \zeta^{n-1} \right) \notag \\
 	 & + \frac{1}{4} e^{2it}\left(\sum_{m=1}^{\infty} (-1)^m i^m J_m(\rho) \zeta^{-m} \right)\left(\sum_{n=1}^{\infty} n (-1)^n i^n J_n(\rho) \zeta^{n-1} \right) .
\end{align}
The residue \eqref{eq:res} picks out the terms with $m=n$.  This reduces the double infinite sums to single infinite sums.  The result is
\begin{align}
\gamma(t,\rho) &= 2\res_{\zeta = 0} \left(  m_- \frac{d}{d\zeta}  m_+ \right) \notag \\
			&= \frac{1}{2} e^{-2it}\sum_{n=1}^{\infty}n(-1)^nJ_n(\rho)^2
	+  \sum_{n=1}^{\infty} n J_n(\rho)^2
	 + \frac{1}{2}  e^{2it} \sum_{n=1}^{\infty} n (-1)^n J_n(\rho)^2  \label{eq:gseries}\,.
\end{align}
To wrap up, we need the identities (see Appendix \ref{sec:app})
\begin{align}
&\sum_{n=1}^{\infty}n(-1)^nJ_n(\rho)^2 = -\frac{1}{2} \rho J_0(\rho) J_1(\rho) \,, \label{eq:bessel1}\\
&\sum_{n=1}^{\infty}n J_n(\rho)^2 = \frac{1}{2}\rho^2 J_0(\rho)^2 + \frac{1}{2}\rho^2 J_1(\rho)^2
	-\frac{1}{2} \rho J_0(\rho) J_1(\rho) \,. \label{eq:bessel2}
\end{align}
Plugging into \eqref{eq:gseries} gives
\beq
\gamma(t,\rho) =  \frac{1}{2}\rho^2 J_0(\rho)^2 + \frac{1}{2}\rho^2 J_1(\rho)^2
	-\rho J_0(\rho) J_1(\rho) \cos^2t  \,,
\eeq
which precisely matches \eqref{eq:gamma}.   

We thus recover the Einstein--Rosen metric \eqref{eq:metric}--\eqref{eq:gamma} from its monodromy function \eqref{eq:monodromy} using only algebra.  The Einstein--Rosen monodromy function can be obtained from the monodromy function of Minkowski spacetime using the action of the Geroch group  \cite{breitenlohner1987geroch}.

\noindent{\it Acknowledgment}  This research was supported, in part, by the U.S. Department of Energy and the Sivian Fund at the Institute for Advanced Study.

\appendix

\section{Bessel function identities}
\label{sec:app}

Our calculation of $\gamma(t,\rho)$ relied on a pair of identities \eqref{eq:bessel1}--\eqref{eq:bessel2} involving infinite sums of squares of Bessel functions.  We record their proofs below for completeness.  Essentially the same identities appear in Watson's treatise on Bessel functions\footnote{I thank David Chow for pointing this out to me.} \cite{watson1995treatise}.

The first identity \eqref{eq:bessel1} is
\beq
\sum_{n=1}^{\infty}n(-1)^nJ_n(\rho)^2 = -\frac{1}{2} \rho J_0(\rho) J_1(\rho) \,.
\eeq
To prove this, use the recurrence relation $nJ_n(\rho) = \frac{\rho}{2}(J_{n-1}(\rho)+J_{n+1}(\rho))$ to get
\begin{align}
\sum_{n=1}^{\infty}n(-1)^nJ_n(\rho)^2 
	&= \frac{\rho}{2}\sum_{n=1}^{\infty}(-1)^n J_n(\rho) (J_{n-1}(\rho)+J_{n+1}(\rho) ) \notag \\
	&= \frac{\rho}{2}\sum_{n=1}^{\infty}(-1)^n J_{n-1}(\rho) J_n(\rho)
	+ \frac{\rho}{2}\sum_{n=1}^{\infty}(-1)^n  J_n(\rho) J_{n+1}(\rho) \,. \label{eq:piece1}
\end{align}
Shift the index on the first term to obtain
\beq
\frac{\rho}{2}\sum_{n=1}^{\infty}(-1)^n J_{n-1}(\rho) J_n(\rho) 
	= - \frac{\rho}{2} J_0(\rho)J_1(\rho) - \frac{\rho}{2}\sum_{n=1}^{\infty}(-1)^n  J_n(\rho) J_{n+1}(\rho) \,. \label{eq:piece2}
\eeq
Combine \eqref{eq:piece1} and \eqref{eq:piece2} to get the desired identity,
\beq
\sum_{n=1}^{\infty}n(-1)^nJ_n(\rho)^2 = -\frac{1}{2} \rho J_0(\rho) J_1(\rho) \,.
\eeq

The second identity \eqref{eq:bessel2} is
\beq\label{eq:id2}
\sum_{n=1}^{\infty}n J_n(\rho)^2 = \frac{1}{2}\rho^2 J_0(\rho)^2 + \frac{1}{2}\rho^2 J_1(\rho)^2
	-\frac{1}{2} \rho J_0(\rho) J_1(\rho) \,. 
\eeq
The easiest way to check that this is correct is to check that both sides have the same derivative with respect to $\rho$, and to check that they have the same value at $\rho=0$.  Clearly both sides vanish at $\rho=0$, so it is enough to check that they have the same derivative.  The derivative of the rhs is 
\beq\label{eq:drhs}
\frac{\rho}{2} J_0(\rho)^2 + \frac{\rho}{2} J_1(\rho)^2 \,.
\eeq
The derivative of the lhs is
\beq
\frac{\partial}{\partial \rho} \sum_{n=1}^{\infty}n J_n(\rho)^2 
	=2  \sum_{n=1}^{\infty}n J_n(\rho) \frac{\partial}{\partial \rho} J_n(\rho) \,.
\eeq
Use the recurrence relations, $nJ_n(\rho) = \frac{\rho}{2}(J_{n-1}(\rho)+J_{n+1}(\rho))$ and $\frac{\partial}{\partial \rho} J_n(\rho) = \frac{1}{2}(J_{n-1}(\rho) - J_{n+1}(\rho))$, to get
\begin{align}
\frac{\partial}{\partial \rho} \sum_{n=1}^{\infty}n J_n(\rho)^2 
	&=\frac{\rho}{2}  \sum_{n=1}^{\infty}(J_{n-1}(\rho)+J_{n+1}(\rho))(J_{n-1}(\rho) - J_{n+1}(\rho))  \notag \\
	&=\frac{\rho}{2}  \sum_{n=1}^{\infty} J_{n-1}(\rho)^2 - \frac{\rho}{2}  \sum_{n=1}^{\infty}  J_{n+1}(\rho)^2  \notag \\
	&=\frac{\rho}{2} J_0(\rho)^2 + \frac{\rho}{2} J_1(\rho)^2  \,.
\end{align}
This is the same as \eqref{eq:drhs}, so we are done.

\bibliographystyle{jhep}
\bibliography{mybib}

\end{document}